\documentclass[10pt,letterpaper]{article}
\usepackage{opex3}

\usepackage{verbatim}

\def\aap{A\&A}%

\def\pasp{PASP}%

\begin{document}

\title{Pupil stabilization for SPHERE's extreme AO and high performance coronagraph system.}

\author{Guillaume Montagnier,$^{1,3}$ Thierry Fusco,$^2$ Jean-Luc Beuzit,$^1$
David Mouillet,$^1$ Julien Charton,$^1$ and Laurent Jocou$^1$}

\address{$^1$LAOG: Laboratoire d'Astrophysique de Grenoble, BP 53,\\
F-38041 Grenoble Cedex 9, France}
\address{$^2$ONERA, BP72, 29 avenue de la Division Leclerc, \\
F-92322 Chatillon Cedex, France}
\address{$^3$Observatoire de Gen\`eve, 51 chemin des Maillettes, \\
1290 Sauverny, Switzerland}

\email{Guillaume.Montagnier@obs.ujf-grenoble.fr} 



\begin{abstract}
We propose a new concept of pupil motion sensor for astronomical
adaptive optics systems and present experimental results obtained
during the first laboratory validation of this concept. Pupil motion
is an important issue in the case of extreme adaptive optics, high
contrast systems, such as the proposed Planet Finder instruments for
the ESO and Gemini 8-meter telescopes. Such high contrast imaging
instruments will definitively require pupil stabilization to minimize
the effect of quasi-static aberrations.
The concept for pupil stabilization we propose uses the flux information
from the AO system wave-front sensor to drive in closed loop a pupil
tip-tilt mirror located in a focal plane.
A laboratory experiment validates this concept and demonstrates its
interest for high contrast imaging instrument.
\end{abstract}

\ocis{010.1080, 220.4830} 




\section{Introduction}

Nowadays, the field of high contrast imaging in astronomy is in great progress.
Indeed, most of the large telescopes in operation are
equipped with adaptive optics systems (NACO on the European Southern
Observatory (ESO) Very Large Telescope (VLT) \cite{rousset2003}, Altair on
Gemini North \cite{stoesz2004}, Keck-AO on the Keck II telescope
\cite{wizinowich2006} for example. Stellar coronagraphy is also being used in
combination with some of these systems \cite{boccaletti2004}.

Some astrophysical programmes are still out of reach with the existing
instruments, due to the relatively limited performance of these
systems in terms of the achieved contrast. More specifically,
extra-solar planets cannot be directly detected with the current
systems, except for a few recently confirmed planetary mass companions
to young brown dwarfs \cite{chauvin2005a, chauvin2005b, neuhauser2005}.
But these very favorable cases, thanks to the young
age and relatively low mass of the host stars not very demanding in terms
of contrast requirements, are not representative of the vast
majority of the envisioned targets.

 The very high contrast required for direct detection of extra-solar
 planets around most of the target stars, is an extremely challenging
 goal, for
 which new and dedicated instruments should be developped.

 In particular, such a new instrument, called SPHERE
 \cite{beuzit2006, dohlen2006} is currently being developped to equip the ESO VLT by
 mid of 2010. SPHERE will combine eXtreme Adaptive
 Optics (XAO) \cite{fusco2006}, high performance coronagraphy
 \cite{boccaletti2004} and spectroscopic and polarimetric differential imaging
 techniques \cite{racine1999, berton2006, schmid2005}.

For very high contrast observations, SPHERE system performance analysis has
shown that
the system pupil stability is one of the main limitations. The new
generation of instruments will therefore need a way to very accurately
control the stability of the pupil, from the telescope mirrors to the
science detector, especially up to the coronagraph.

We present here our study on pupil stabilization for XAO systems,
performed as part of the SPHERE phase A study. We first
discuss the impact of pupil stability for extra-solar planets
detection in section~\ref{sec:planets}. 
In section~\ref{sec:design} 
we present the
principle of the pupil motion control and in section~\ref{sec:error} we
discuss the possible sources of errors in the measurement of the pupil
shift. Finally, in section~\ref{sec:experiment} we present the laboratory
experiment which allowed to fully validate the proposed concept.

\section{Impact of pupil motion on extra-solar planet detection}

\label{sec:planets}

\begin{figure}[htbp]
\begin{center}
\begin{tabular}{ccccc}
\includegraphics[width=0.2\linewidth]{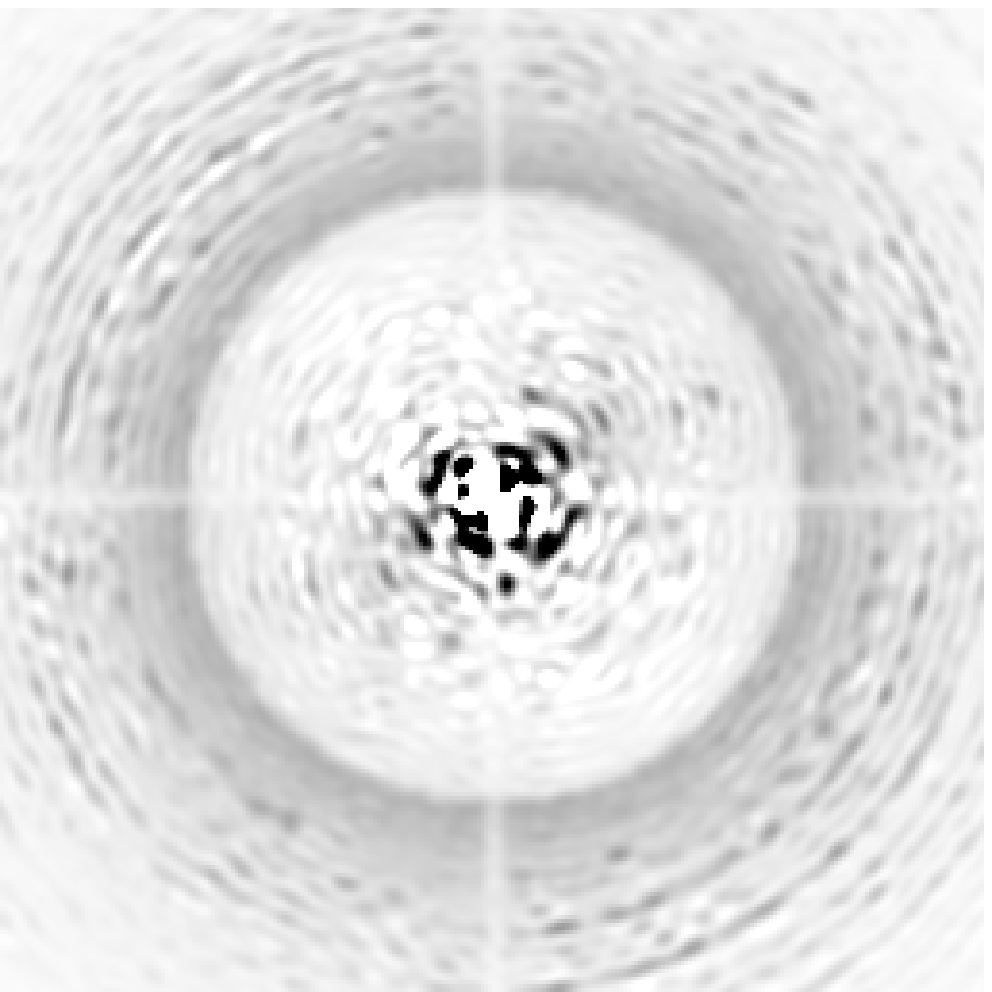}&\hspace{0.0\linewidth}  & \includegraphics[width=0.2\linewidth]{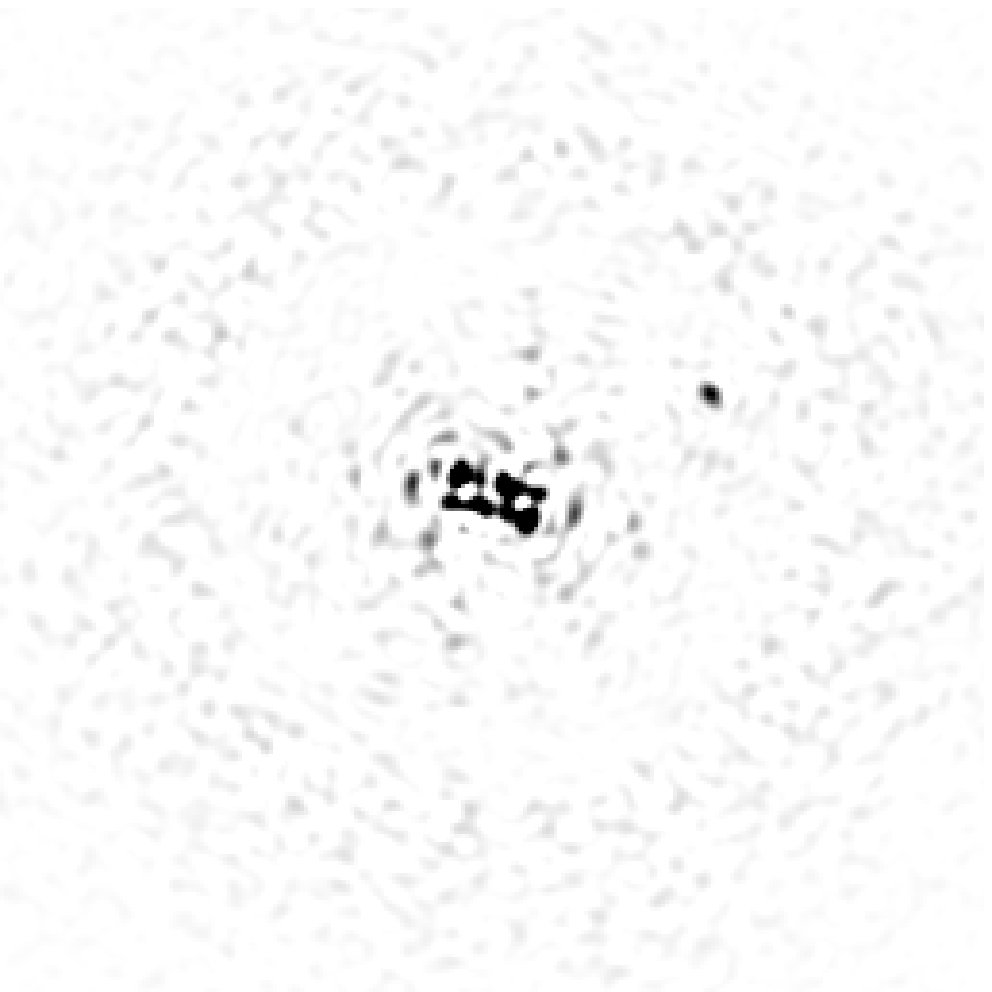}& \includegraphics[width=0.2\linewidth]{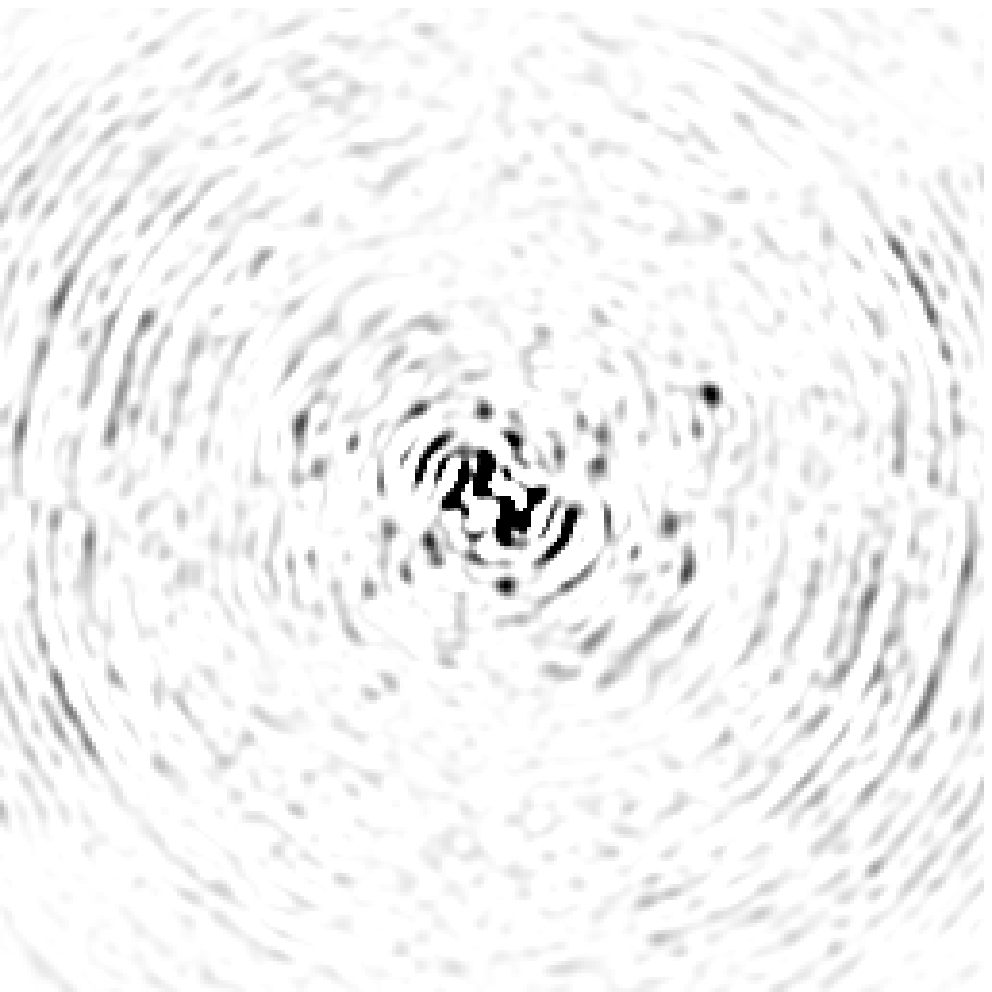}& \includegraphics[width=0.2\linewidth]{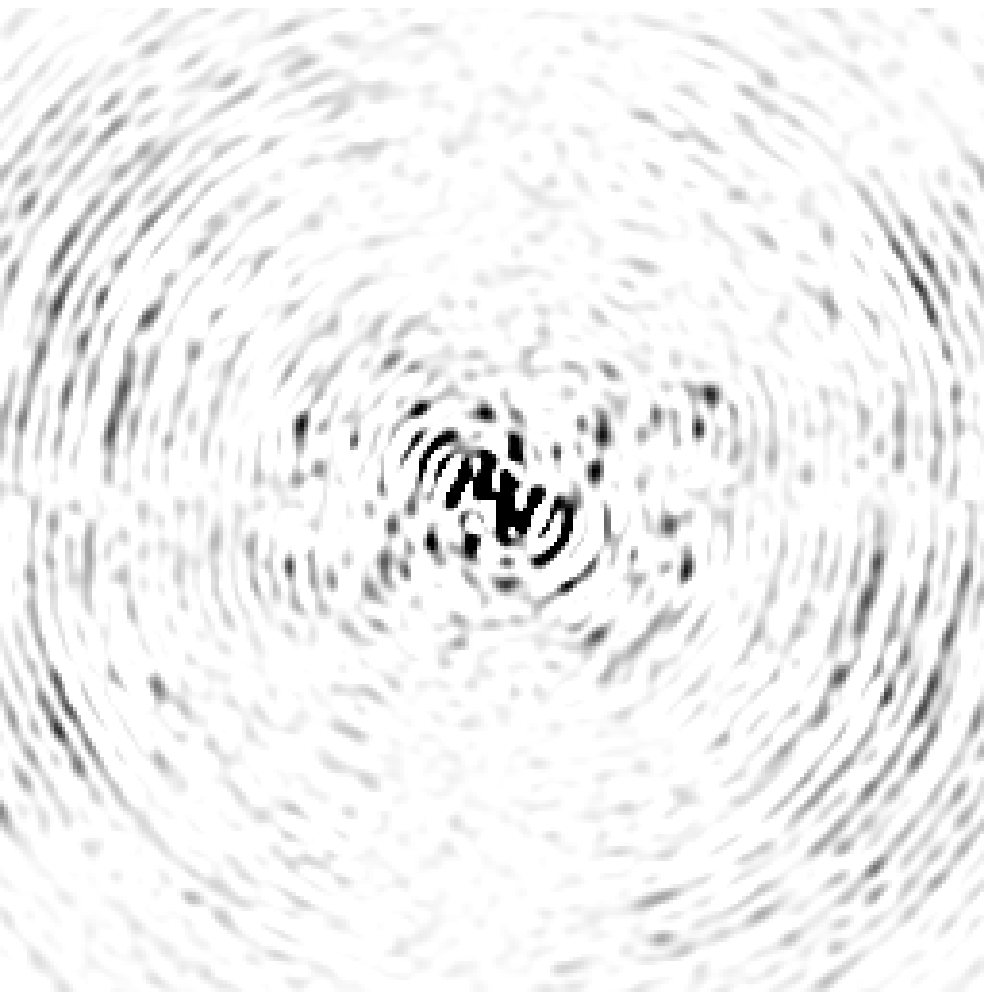}\\
 differential  image   &  \hspace{0.0\linewidth}    & motionless pupil & pupil
 motion = 0.6\% & pupil motion = 1.2\% \\
 & \hspace{0.0\linewidth}  &  \multicolumn{3}{c}{Differential
   image + calibration on reference star}\\
 $I(\lambda_1)-I(\lambda_2) $& \hspace{0.0\linewidth}  &  \multicolumn{3}{c}{$\left
    [I(\lambda_1)-I(\lambda_2)\right ] - \left [Ref(\lambda_1)-Ref(\lambda_2)\right ] $}\\
\end{tabular}
\caption{[Left] Differential coronagraphic (4-quadrants) image
($\lambda_1 = 1.56 \mu m$, $\lambda_2 = 1.59 \mu
  m$), [Right]  differential coronagraphic image + reference subtraction:
  Pupil shift between object and reference star = 0, 0.6 and 1.2 \% of the full
  pupil. The companion ($\Delta_m = 15$, separation = 0.6 arcsec) is really distinguishable from
  residual fixed speckles for a fixed pupil. \label{fig:effect_shift}}
\end{center}
\end{figure}

 Extra-solar planet detection using ground based telescopes requires an
 XAO system to compensate for atmosperic turbulence and system
 aberrations, a coronagraphic device to cancel out the flux coming from
 the central star and a differential imaging technique (for instance a
 subtraction of two images obtained at two nearby wavelengths) to
 cancel out the residual un-corrected PSF pattern \cite{racine1999}. 
 Limitations to this approach are mainly due to the differential
 aberrations in the two image channels. These differential aberrations
 mix up with fixed uncorrected aberrations (coming from
 non common path errors), leading to residual fixed
 patterns after image subtraction 
 (see our simulation on figure~\ref{fig:effect_shift}).
 Such optical wavefront errors come
 from instrumental aberrations. Even if they are small, their temporal
 behavior is problematic due to system temporal evolution (in particular
 the pupil motion during the observation sequence): 
 they are not perfectly constant and are
 therefore difficult to calibrate; their variability is however much
 too slow to be averaged with time, as with turbulence residuals.

 In order to calibrate the residual pattern in differential imaging and
 thus to reach SPHERE required contrast of less than $10^{-6}$ in H band 
 at 0.5 arcseconds,
 observations have to be obtained using a reference star. To be
 efficient, this calibration has to be performed with the whole system
 as stable as possible.

 In particular, the telescope pupil has to remain motionless during the
 whole observing process, typically 1 or 2 hours. When located at the
 Nasmyth focus of a telescope, as SPHERE instrument will be, this
 stability requirement implies the
 use of a pupil de-rotator and a pupil re-centering device. It has been
 shown on simulations that a pupil shift of 1\% of the pupil diameter
 or a pupil rotation of 1 degree will reduce by a factor of 1.5 to 2
 (for typical observing conditions at the VLT) the detection capability
 of a coronagraphic differential imaging system.

 For example, the behavior of the telescope pupil on NACO gives an
 order of magnitude of the pupil shift: when NACO is rotated by 180
 degrees the pupil shift reaches 2\% of the pupil diameter. This pupil
 shift is due to rotation axis misalignment, flexures of the telescope
 structure, M3 mirror misalignment, Nasmyth platform deformation, as
 well as pupil rotation errors in the case of SPHERE (non-perfect derotation
 system at the entrance of the SPHERE bench).

 In terms of dynamical evolution, a first estimate leads to a pupil
 shift of 0.1\% over 15 minutes. 
 Simulations for the SPHERE have shown that the pupil
 stability in translation in any pupil plane must be better than 0.2\%
 (goal 0.1\%) of the pupil diameter in order to allow the detection of
 self-luminous young planets around solar type stars within 50
 parsecs from the sun. As a typical observing sequence (science
 exposure followed by calibration exposure on a reference star) lasts
 for about an hour, a closed loop system has to be implemented in order
 to fulfil this tight requirement. This device will be composed of a 
 pupil sensor and a pupil motion corrector.

\section{Design of a closed-loop pupil shift corrector}

\label{sec:design}

 The basic idea is to measure the pupil position
 without adding any dedicated sensor to avoid
 splitting the light beam and therefore wasting precious photons. 
 Another goal was to try to keep the system as simple as possible
 in order to reduce its impact on the overall SPHERE instrument.
 Considering these two requirements we have chosen to use
 the existing Shack-Hartmann Wave Front
 Sensor (SH WFS) of the main adaptive optics loop. The only additionnal
 device will be a pupil tip-tilt mirror (PTTM) located close to the input
 focal plane of the instrument in order to correct for any measured pupil
 motion.

 \subsection{Principle of the sensor}
 
\begin{figure}
\begin{center}
\includegraphics[width=\linewidth]{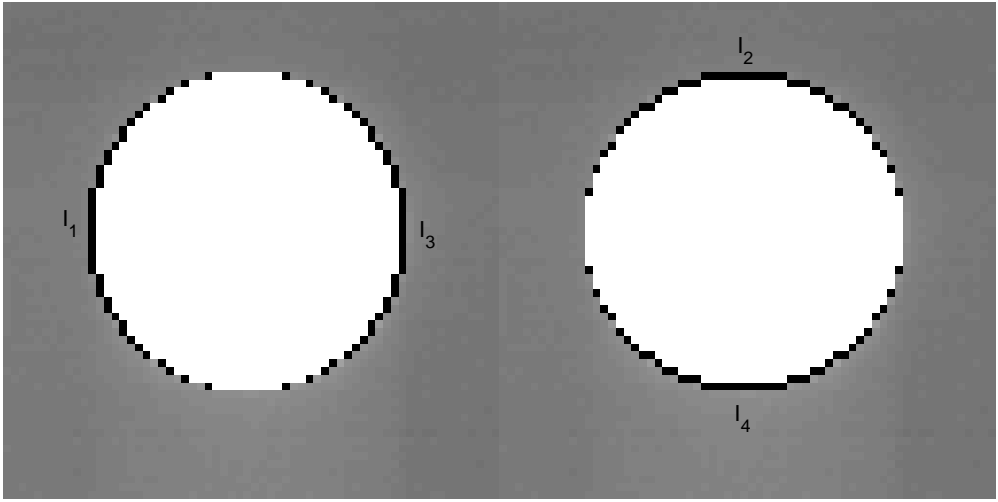}
\end{center}
\caption{\label{fig:sub-aperture_pattern} Ring-shaped pattern for
sub-apertures.}
\end{figure}

 The sensing principle consists in measuring the integrated flux in 4
 computation areas that include sub-apertures located at the edges of 
 the pupil (hereafter areas are labeled 1, 2, 3 and 4, 
 see figure~\ref{fig:sub-aperture_pattern}).

 From the four integrated flux values ($I_1$, $I_2$, $I_3$ and $I_4$),
 we compute both $I_x$ and $I_y$:

 \begin{eqnarray}
   I_x=\frac{I_1-I_3}{I_1+I_3} \label{eq:shift-value}\\
   I_y=\frac{I_2-I_4}{I_2+I_4}
 \end{eqnarray}

 which give the information on pupil shifts along the x and y axes
 respectively.

 For practical purpose, all the sub-apertures included in the four
 areas on figure~\ref{fig:sub-aperture_pattern} will not be used.
 Only sub-apertures close to the x-axis
 (respectively y-axis) will be included in computation areas to measure the 
 shift along the x-axis (respectively y-axis) because the farther the 
 sub-apertures are from the axis the larger is the noise in the $I_i$ signal.
 For example, the flux in a sub-aperture located on the y-axis does not bring
 any information on the shift along the x-axis, it only brings noise.
 Therefore, for analytical developpement, we will consider the pessimistic
 case where we use only sub-apertures close to the x and y axis.
 We will discuss the general case in the experimental section
 (see section~\ref{sec:experiment}).

 \begin{figure}
   \begin{center}
     \includegraphics[width=\linewidth]{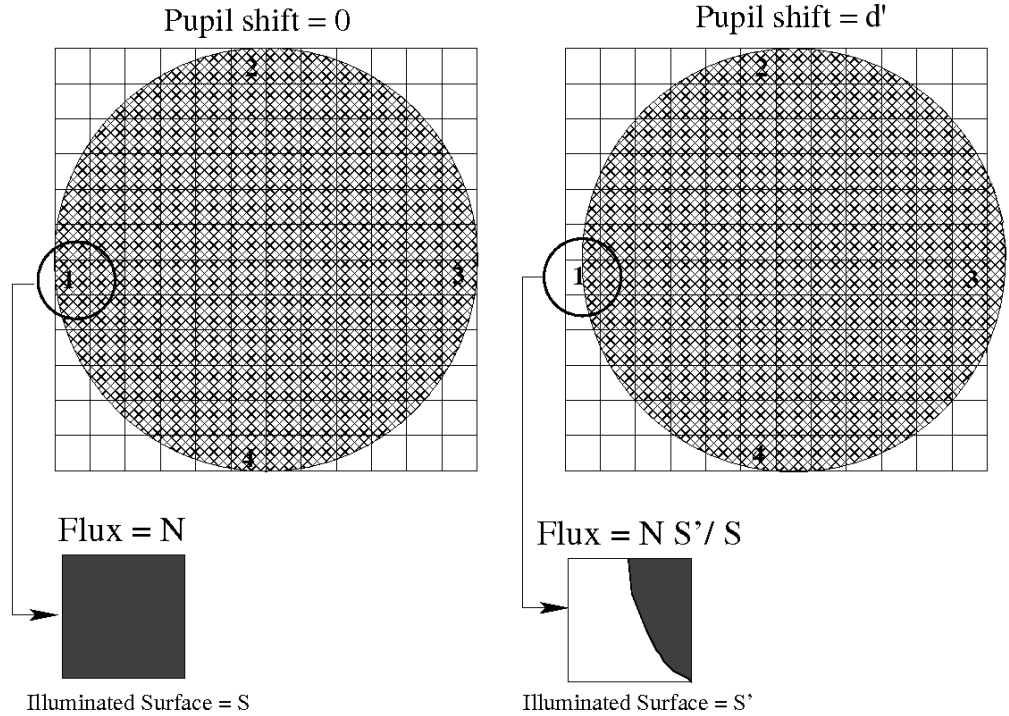}
   \end{center}
   \caption{\label{fig:principle} Schematic representation of the
     pupil shift measurement using a SH WFS.}
 \end{figure}

\begin{figure}
   \begin{center}
     \includegraphics[width=\linewidth]{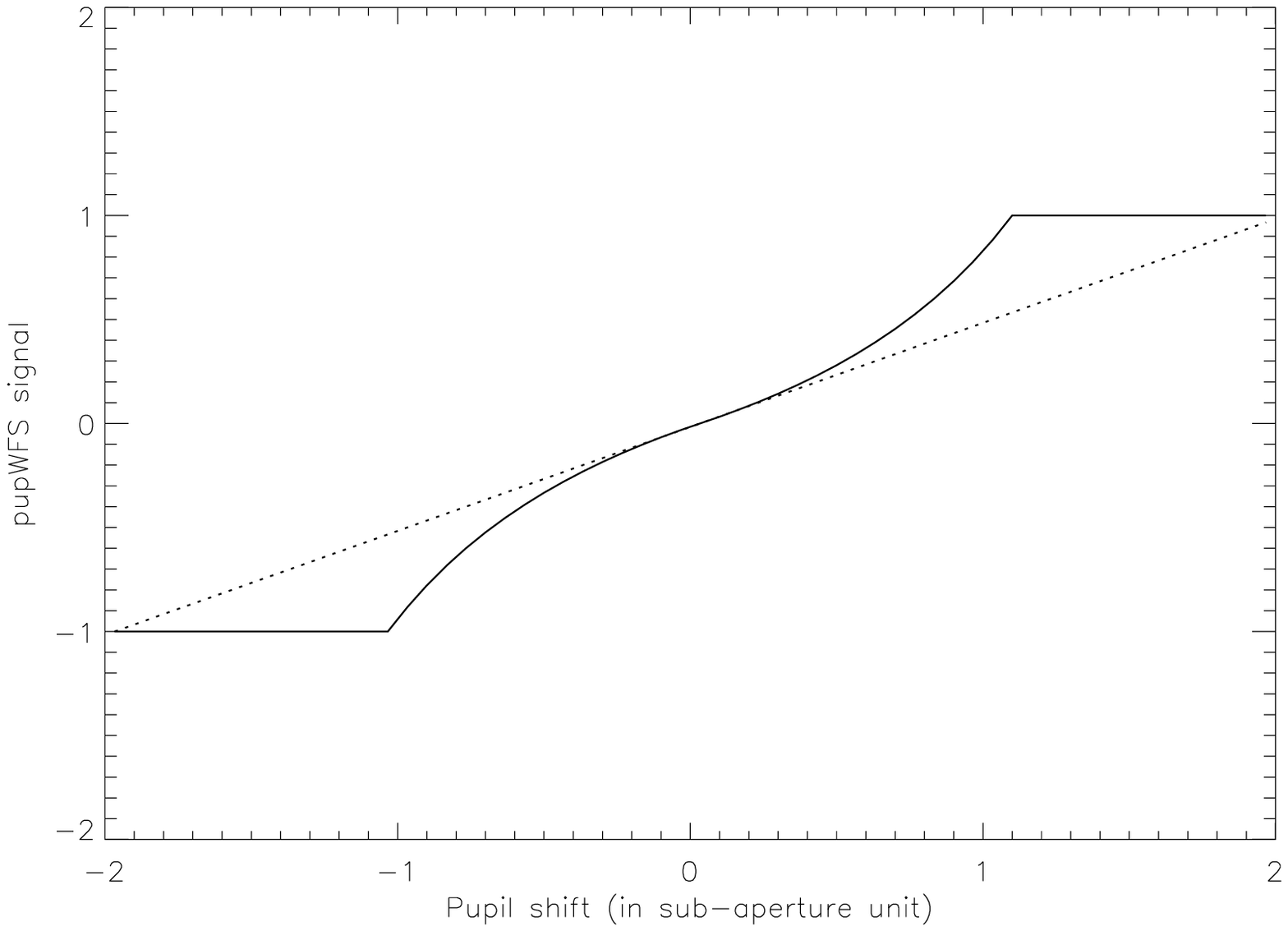}
   \end{center}
   \caption{\label{fig:shift}$I_x$ as a function of pupil shift (in
     sub-aperture unit).}
 \end{figure}

 The measurement principle is shown on
 figure~\ref{fig:principle} with only one sub-aperture per area.
 Each square in the figure represents one of the
 WFS sub-apertures. Hereafter, we will use the representation
 of this figure to develop the equations.
 In this case,  $I_i=\textrm{N} \cdot \textrm{S'} / \textrm{S}$ 
 (S and S' are defined on figure~\ref{fig:principle}).
 Assuming  that the edge of
 the telescope is a straight line on the computation area used to
 measure the pupil motion and that only sub-apertures close to the 
 axis are used , the pupil shift amplitude $d'$ along the
 $x$ axis is directly linked to $I_x$ by:

 \begin{equation}
   \label{eq:approximation}
   \left|{I_x}\right|=\frac{d'}{2 \cdot d}
 \end{equation}

 where d is the sub-aperture size and $d \gg d'$ (with this approximation
 $\textrm{I}_i \propto (\textrm{d}-\textrm{d'}) \cdot l$ where
 l is the width of the computation area). 
 The same applies to
 the y axis. The sign of $I_x$ (respectively $I_y$) gives the
 direction of the shift. We present in figure~\ref{fig:shift} the $I_x$
 signal as a function of the pupil shift. For comparison purposes,
 the theoretical expression of the signal (obtained for $d \gg d'$)
 is also plotted. One can note that linearity is maintained for
 shifts lower than typically $25 \%$ of a sub-aperture size.

 \subsection{Measurement of the integrated flux}

 The measurement of the integrated flux per sub-aperture must be
 obtained from the Shack-Hartmann spots after flat-field correction and
 background subtraction.
 In order to increase the measurement accuracy, the signal of several
 computation area located at the pupil edge can be combined to obtain the final $I_i$ values,
 including areas next to the central obscuration in the case of an
 astronomical reflective telescope.

 \subsection{principle of the correction}

 We first need to measure the flux received by each sub-aperture using
 a calibration source (located at the entrance of the adaptive optics
 system) to determine reference values. Using these reference values,
 the pupil is roughly centered "by hand". Then, an interaction matrix
 is acquired using the tip-tilt mirror located in a focal
 plane. Next, a command matrix is computed by direct inversion of the
 interaction matrix. Finally, when the pupil motion correction system
 is running, a closed loop measurement of $I_1$, $I_2$, $I_3$ and $I_4$
 allows to compute the corrections which are sent to the control loop
 of the pupil tip-tilt mirror.

\section{Performance study}

\label{sec:error}

 Before performing an experimental validation of the concept, an
 analytical study has been conducted. In this study, the different
 sources of measurement error are dicussed, as well as the impact of
 the atmospheric turbulence.

 \subsection{Measurement error}
       
 \subsubsection{Photon noise\\}

 \label{sec:photon_noise}

 From equation~\ref{eq:shift-value}, the noise variance on the
 measurement signal $I_x$ is:

 \begin{equation}
 \sigma^{2}_{I_x}=\left\langle \left( \frac{I_1-I_3}{I_1+I_3}
 \right)^2 \right\rangle - \left( \left\langle
 \frac{I_1-I_3}{I_1+I_3} \right\rangle \right)^2
 \label{eq:variance}
 \end{equation}

 Assuming that the noise fluctuations of $I_1+I_3$ are negligible and
 that $I_1 + I_3= N+N'$ with $N$ being the number of photons in a fully
 illuminated area and $N'$ the flux in the truncated area ($N' \simeq
 N(1-d'/d)$ if $d' \ll d$), equation~\ref{eq:variance} becomes:

 \begin{equation}
 \sigma^{2}_{I_x}=\frac{\langle I_1^2 \rangle - \langle I_1
 \rangle^2 + \langle I_3^2 \rangle -\langle I_3 \rangle^2 -2\left(
 \langle I_1\cdot I_3 \rangle - \langle I_1 \rangle \langle I_3
 \rangle \right)}{(N+N')^2} \label{eq:variance:bis}
 \end{equation}

 Then assuming that $I_1$ and $I_3$ follow photon noise statistics
 and that $I_1$ and $I_3$ noises are decorrelated, it comes:

 \begin{equation}
 \sigma^{2}_{I_x}=\frac{N+N'}{(N+N')^2}
 \end{equation}

 and if $d \ll d'$, $N \simeq N'$, then equation~\ref{eq:variance:bis}
 finally leads to:

 \begin{equation}
 \sigma^{2}_{I_x}\simeq\frac{1}{2N} \label{eq:variance_photon}
 \end{equation}

\begin{figure}
\begin{center}
\includegraphics[width=\linewidth]{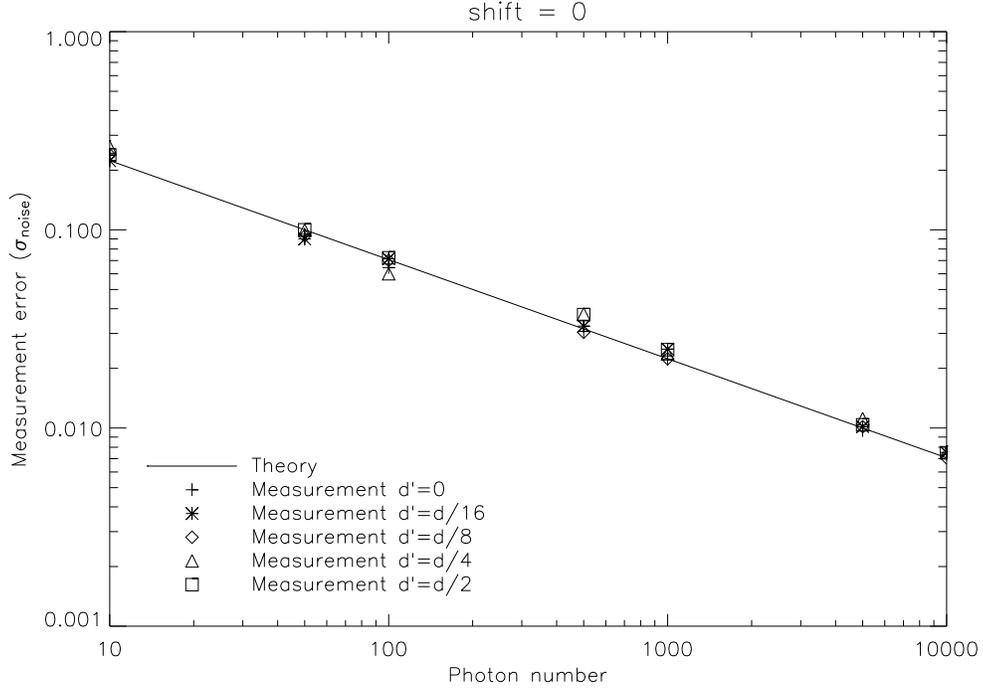}
\end{center}
\caption{\label{fig:inf_noise_phot} Photon noise influence on
pupil sensor measurement.}
\end{figure}

 Figure~\ref{fig:inf_noise_phot} confirms the hypothesis made to obtain
 equation~\ref{eq:variance_photon}. A $15\times 15$ Shack-Hartmann has
 been simulated (only considering its geometry)
 by a flat incoming wavefront and a varying number of
 photons per computation area and per frame. The simulation shows that
 the approximations used in equation~\ref{eq:variance_photon} are
 still valid up to $d'=d/2$.

 \subsubsection{Detector noise\\}

 Following the same development as in section~\ref{sec:photon_noise}, the
 noise variance in the case of detector noise can be written as:

 \begin{equation}
 \sigma^{2}_{I_x}\simeq \frac{2\cdot \sigma_{detec}^2}{(N+N')^2}
 \end{equation}

 where $\sigma_{detec}$ is equal to $N_{pix}\times RON^2$ with
 $N_{pix}$ the number of pixels per computation area and $RON$ the
 detector Read-Out Noise. Again assuming $d'\ll d$, it comes:

 \begin{equation}
 \sigma^{2}_{I_x}\simeq \frac{\sigma_{detec}^2}{2N^2}
 \label{eq:variance_detector}
 \end{equation}

 \subsubsection{Noise effect on pupil sensor measurement\\}

 Let us consider that the pupil sensor works in closed-loop (in term of
 pupil centering). In that case, the assumption that $d \ll d'$ is well
 verified. Then, the effects of noise (detector noise and photon noise
 can be summarized in one global equation (from 
 equations~\ref{eq:variance_photon} and~\ref{eq:variance_detector}):

 \begin{equation}
 \sigma^{2}_{I_x}\simeq \frac{N+\sigma_{detec}^2}{2N^2}
 \label{eq:variance_noise}
 \end{equation}

 It is clear, from equation~\ref{eq:variance_noise} that as soon as
 $N\gg\sigma_{detec}^2$, the dominant effect is the photon noise.

 We see from equation~\ref{eq:variance_noise} that measurement
 sensitivity and photons number per sub-aperture increase accordingly.
 For a given flux on the camera, another way to increase the
 sensitivity is to use a greater number of sub-apertures to measure the
 integrated flux in the 4 computation areas.

\subsubsection{Impact of integration time on measurement  noise \\}

Assuming the following parameters

\begin{figure}[htbp]
 \centering \includegraphics[width=0.8\linewidth]{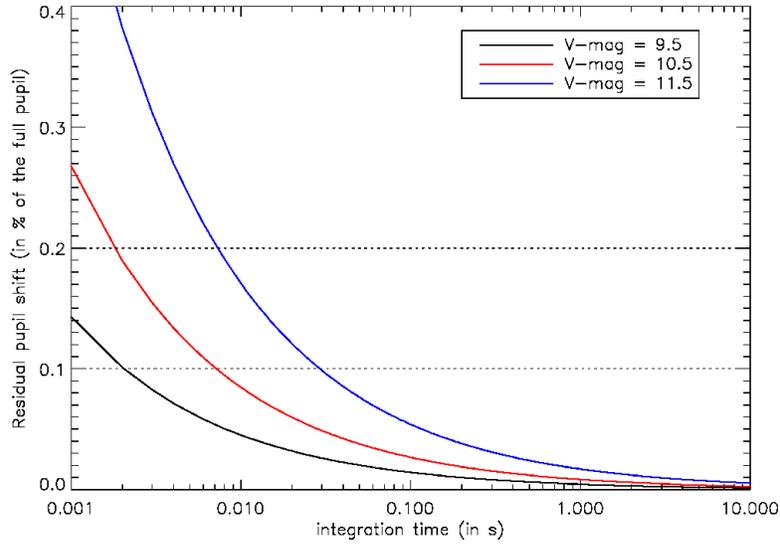}
 \caption{Evolution of the residual pupil motion as a function of the
   integration time. A closed loop measurement is considered, meaning
   that the noise is filtered by the loop (as a function of its
   bandwidth: $BW = F_{samp}/6)$.
 \label{fig:in_int}}
 \end{figure}

\begin{itemize}
\item   1e- RON CCD
\item   1 kHz sampling frequency
\item   6x6 pixels per sub-aperture
\item   Three GS magnitudes (M0 spectral type) : 9.5 (130 photons/sub-aperture/frame),
10.5 (50 photons/sub-aperture/frame), and 11.5 (20 photons/sub-aperture/frame)
\end{itemize}

one can deduce the pupil motion sensor performance as a function of
 the integration time. Considering that the typical evolution of the
 sub-aperture position is 0.1\% every 15 minutes (for alt-azimuthal
 telescopes, it will of course depend on the zenith angle), the typical
 exposure time of the pupil motion sensor has to be smaller than a few
 minutes. In closed loop operation, the system bandwidth is typically 6
 to 15 times smaller that the sampling frequency depending on the loop
 scheme and the overall delay. In our case, the time scale is pretty 
 large (a few seconds to a minute) which will probably dramatically reduce
 the computational delay (with respect to integration time itself). It
 therefore seems reasonable to consider a system bandwith equal to
 1/6th of the sampling frequency.
 The results in Figure \ref{fig:in_int} show that the noise
 measurement will not be a limitation in terms of system
 accuracy. In any case, one can increase the number of
 sub-apertures (for instance all the
 sub-apertures located at the edge of the pupil) in order to improve the
 signal to noise ratio.


   \subsection{Impact of atmospheric turbulence}
 Another contribution to the flux fluctuation in each sub-aperture is
 the atmospheric effects which can be decomposed in two main
 contributors:
 \begin{itemize}
 \item the scintillation effects which induce amplitude fluctuation of the
   electromagnetic field during the propagation trough the turbulence and thus
   intensity fluctuation at the level of the focal plane of  each sub-aperture
 \item the speckle pattern in each sub-aperture PSF which randomly evolves and
   induces intensity fluctuations because of the finite size of the sub-aperture
   FoV (typically a few arcseconds).
 \end{itemize}
 These two effects are quantified in the following sub-sections.

         \subsubsection{Scintillation effect \\}

 Scintillation effects lead to sub-aperture flux variations and thus can affect
 the pupil motion sensor measurements.
  The typical size of a scintillation pattern (in the
 Rytov approximation, well validated for astronomical observation) is
 given by:
 \begin{equation}
 L_{scint}= \sqrt{\lambda_{wfs} h_{eq}}
 \end{equation}
 where $h_{eq}$ represents an equivalent distance of propagation
 (typicaly a few kilometers for astronomical site).
 $h_{eq}$ is given (in the Rytov approximation) by \cite{clifford1978}:
 \begin{equation}
 h_{eq} = \left (\frac{\int_{0}^{\infty} h^2 C_n^2(h) dh}{\int_{0}^{\infty}  C_n^2(h)
   dh}\right )^{1/2}
 \end{equation}
 and $\lambda_{wfs}$ the WFS wavelength. If a visible WFS is
 considered, $\lambda_{wfs} = 0.7 \mu m$ typically which leads to a
 typical size of the scintillation pattern equal to a few millimiters
 (4.5 millimeters for a equivalent altitude of 3 km).

 In this conditions, one can assume that the scintillation patterns are by far
 smaller than the sub-aperture size (20 cm for the SPHERE system). Hence, the
 flux fluctuation per sub-aperture and per frame is given by the following
 expression \cite{roddier1981}
 \begin{equation}
 \sigma_I = \sqrt{17.36*d^{-7/3}\int_{0}^{\infty} h^2 C_n^2(h) dh}
 \end{equation}
 with $d$ the sub-aperture size and $C_n^2(h)$ the turbulence profile.
 For typical atmospheric condition, $\sigma_I$ is equal to a few percents.

 This value has to be reduced as a function of the integration time.
 It decreases with the  factor  $\sqrt{T/\tau_{scint}}$ where $T$ is the total
 pupil motion sensor integration time (typical 1 min) and $\tau_{scint}$ the typical
 scintillation lifetime. In a first approximation, and assuming a Taylor
 hypothesis
 \begin{equation}
 \tau_{scint}= L_{scint}/v
 \end{equation}
 with $v$ the turbulence wind speed.

 Hence, considering a 10 m/s wind speed and an integration time of
 typically 10 seconds (which is fast enough to correct for pupil
 motion at the required level of accuracy), the scintillation effects
 lead to flux fluctuation smaller than 0.1 \%  i.e. an error on pupil
 position of 0.005 \%.

 Considering the orders of magnitude involved here (especially pupil motion sensor
 integration time), the scintillation effects are assumed to be
 negligible in the global pupil motion sensor error budget.

         \subsubsection{Effect of sub-aperture field of view\\}
\begin{figure}[htbp]
 \begin{center}
 \includegraphics[width=0.8\linewidth]{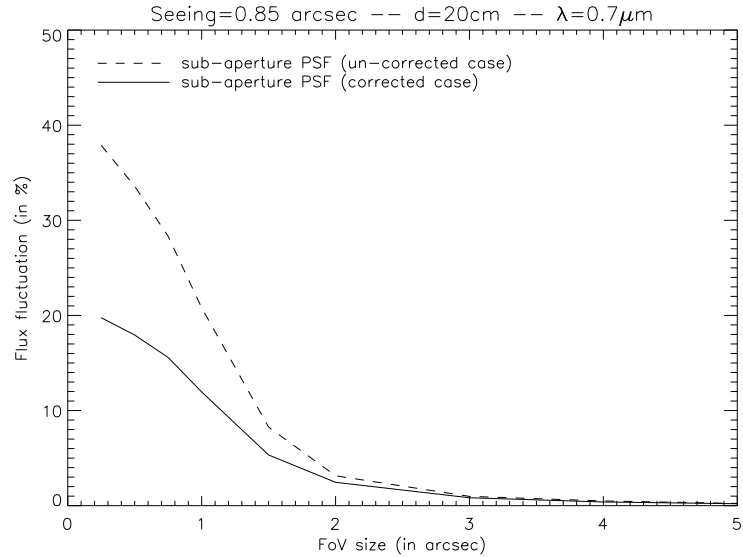}
 \caption{Instantaneous flux fluctuation due to the sub-aperture FoV size \label{fig:inffov}}
 \end{center}
 \end{figure}

 The sub-aperture FoV leads to a flux loss (PSF turbulent wings).
 This flux loss evolves with the residual un-corrected turbulence and
 induces flux variations in the sub-aperture. We plot in Figure
 \ref{fig:inffov} the flux variation in one sub-aperture as a
 function of the FOV size. SPHERE conditions are considered here:
 i.e. a 0.85 arcsec seeing, a 20 cm sub-aperture size and a WFS
 wavelength of 0.7 $\mu m$.

\begin{figure}[htbp]
  \begin{center}
  \includegraphics[width=0.8\linewidth]{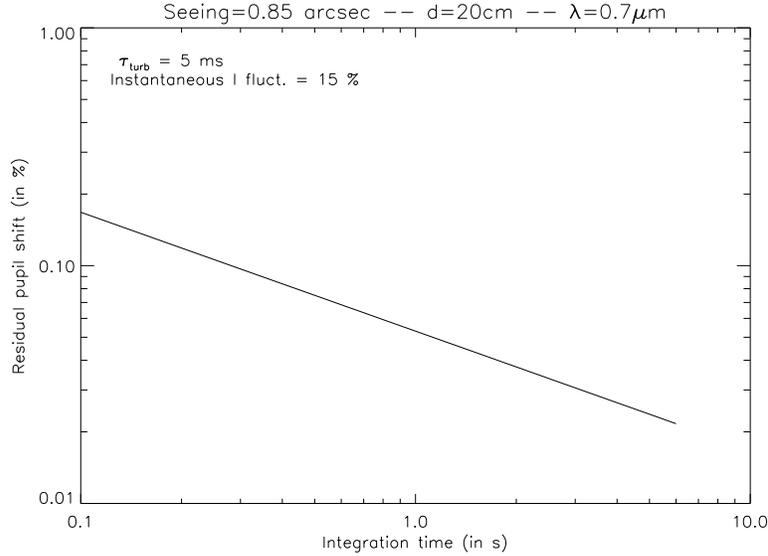}
  \caption{Precision (in \%) on pupil shift measurement as a function of pupil motion sensor
    integration time (only the effects of the finite FoV of the sub-apertures
    are considered here). $\tau_{turb} = 5 ms$, seeing = 0.85 arcsec. \label{fig:evol_with_time}}
  \end{center}
  \end{figure}

  Figure \ref{fig:inffov} shows that for typical FoV (1 to 2 arcseconds) the
  instantaneous flux fluctuations are of the order of 10 to 20 \% which leads to an
  error on pupil position between 0.5 and 1 \%.
 As presented in the previous section for the scintillation case,
 this residual turbulence effect decreases with
 $\sqrt{T/\tau_{turb}}$  where $T$  is the total pupil motion sensor integration time
 (typical 1 min) and $\tau_{turb}$ the typical residual uncorrected
 turbulence evolution (typical a few ms) \cite{roddier1981, fusco2004}. This
 dependency is illustrated in Figure \ref{fig:evol_with_time} where
 the precision on pupil shift estimation is plotted as a function of
 the pupil motion sensor integration time.

  Considering the orders of magnitude involved here (especially pupil motion sensor
  integration time), the previous analysis has shown that the residual uncorrected 
  turbulence effects are
  negligible in the global pupil motion sensor error budget.
  In addition, one can note that the quantification presented here
  is probably pessimistics since the SPHERE system will integrate a spatial
  filter device in front of its wave front sensor \cite {poyneer2004}. This
  device will cancel out all the high spatial frequencies before the wave front 
  sensor measurement. It therefore dramatically reduce the speckle pattern in
  each sub-aperture and leads to very clean and symetrical spots \cite{fusco2005}.

 \section{First experimental validation}

 \label{sec:experiment}
 In order to validate our analytical study, a
  laboratory experiment has been elaborated.

      \subsection{Experimental set-up}

  A test bench has been installed at our laboratory to validate the
  pupil motion sensor concept. It is composed of a white light fiber,
  a pupil diaphragm, a tilt-tilt mirror located in a focal plane and a
  Shack-Hartmann wavefront sensor (HASO 64 from Imagine Optic). The
  Shack-Hartmann wavefront sensor is composed of a $64\times64$
  micro-lens array focusing light on a DALSA CCD camera
  (
  $\textrm{Read Out Noise}=51 \textrm{e-}^2$). We consider only a $40\times40$
  sub-apertures illuminated disc over the $64\times64$
  sub-apertures.

 A first lens is located at a focal distance from a light source to
 obtain a parallel incident beam. The entrance pupil of the
 telescope is defined by a diaphragm, optically conjugated with the
 lenslet plane of the SH WFS. The Tip-Tilt Mirror (TTM) is located
 in a focal plane in order to control the pupil movement. The TTM
 actuators generate pupil shifts in the X and Y directions. \\

      \subsection{Comparison between analytical and experimental results}
      
          \subsubsection{Study of the Pupil Motion Sensor response}

              \paragraph{Choice of the computation areas\\}

              Four computation areas are located at the edge of the pupil.
              The width of each sub-aperture allows a maximum pupil
              shift measurement of $1/40^{th}$ pupil (i.e. $2.5\%$
              of the pupil diameter). A pattern with the maximum of
              sub-apertures per computation areas(41 sub-apertures surrounding the pupil 
              at the most) is defined on
              figure~\ref{fig:sub-aperture_pattern}. Each pixel in the
              figure represents the integrated flux of a SH WFS
              sub-aperture.

             We study in the following paragraphs
              how the pupil measurement evolves with the number of
              useful sub-apertures. One can already guess that,
              since the integrated flux increases with the number of
              sub-apertures, the sensitivity of the measurement will
              increase accordingly as long as the selected sub-apertures
              are close to the corresponding axis 
              (so the equation~\ref{eq:approximation} is valid).

             \paragraph{Response curves\\}

\begin{figure}
\begin{center}
\includegraphics[width=\linewidth]{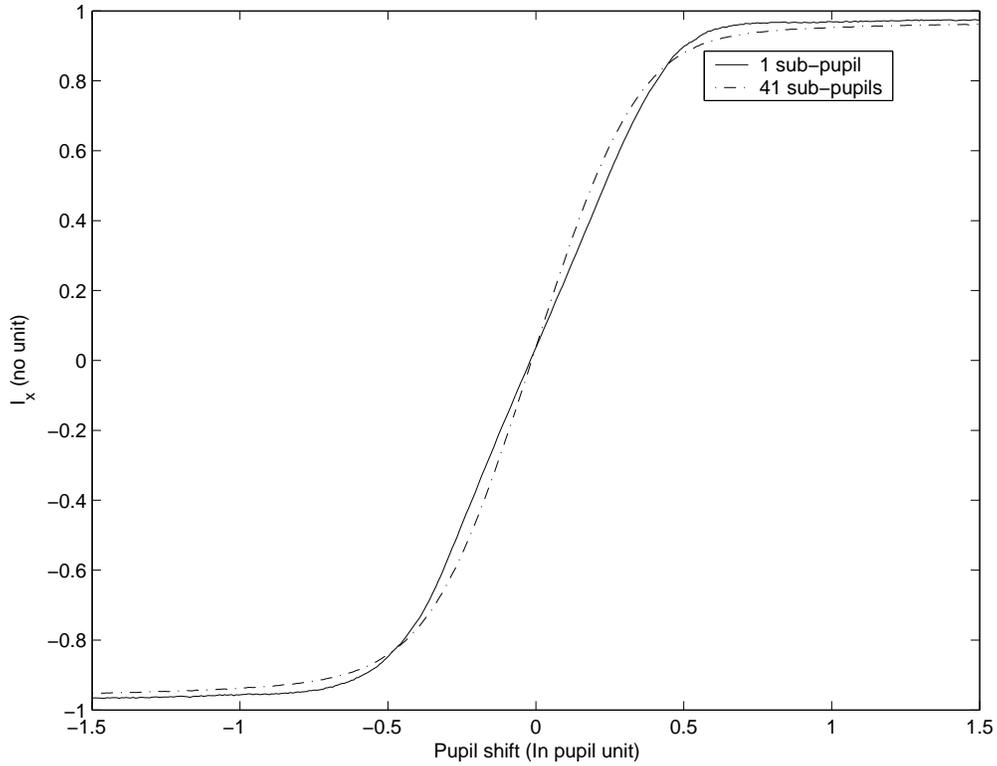}
\end{center}
\caption{\label{fig:response_curves} Response curves for
sub-apertures with 1 and 41 sub-apertures.}
\end{figure}

             We present in figure~\ref{fig:response_curves} the
             experimental signal $I_x$ as a function of pupil shift $S$.  The pupil
             shift is a linear function of the voltage $V_x$ applied to
             the x-actuator of the TTM.
             Two curves are represented, showing two different
             configurations for the computation areas:
             \begin{itemize}
             \item with 4 sub-apertures (one per direction).
             \item with all the  sub-apertures located at the edge and surrounding the
               telescope pupil.
             \end{itemize}

             Figure~\ref{fig:response_curves} shows that the curve slope increases
             when the number of
             sub-apertures increases meaning that  the pupil
             motion sensor sensitivity increases with the number of
             sub-apertures. It is clear that the larger the number of
             sub-apertures (bringing information on the signal of interest)
             the better the measurement accuracy should be.

         \subsubsection{Example of reconstruction\\}

         After the acquisition of an interaction matrix and a computation of a command
         matrix (direct inversion of the interaction matrix), pupil
         shift measurements (i.e. the intensity fluctuation on the
         considered sub-apertures, $I$) and their corresponding  pupil
         shift ($S_{rec}$) are estimated.

\begin{figure}
\begin{center}
\includegraphics[width=\linewidth]{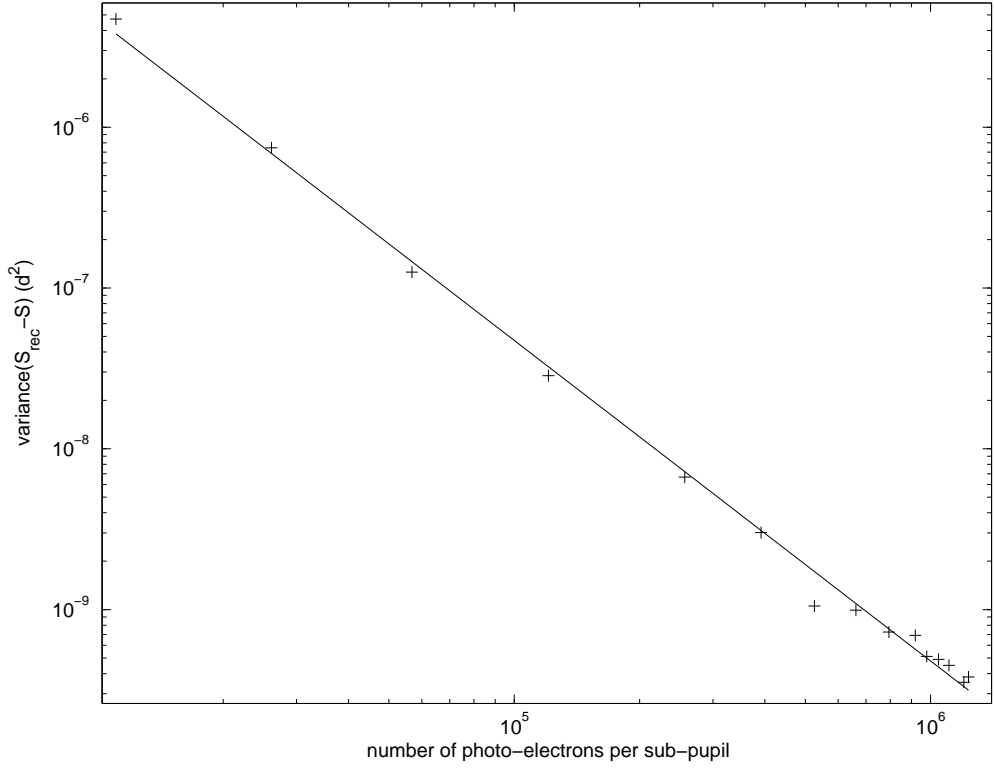} 
\caption{\label{fig:reconstruction_error_fit}Reconstruction error as a
function of the number of photons with one sub-aperture.}
\end{center}
\end{figure}

\begin{figure}
\begin{center}
\includegraphics[width=\linewidth]{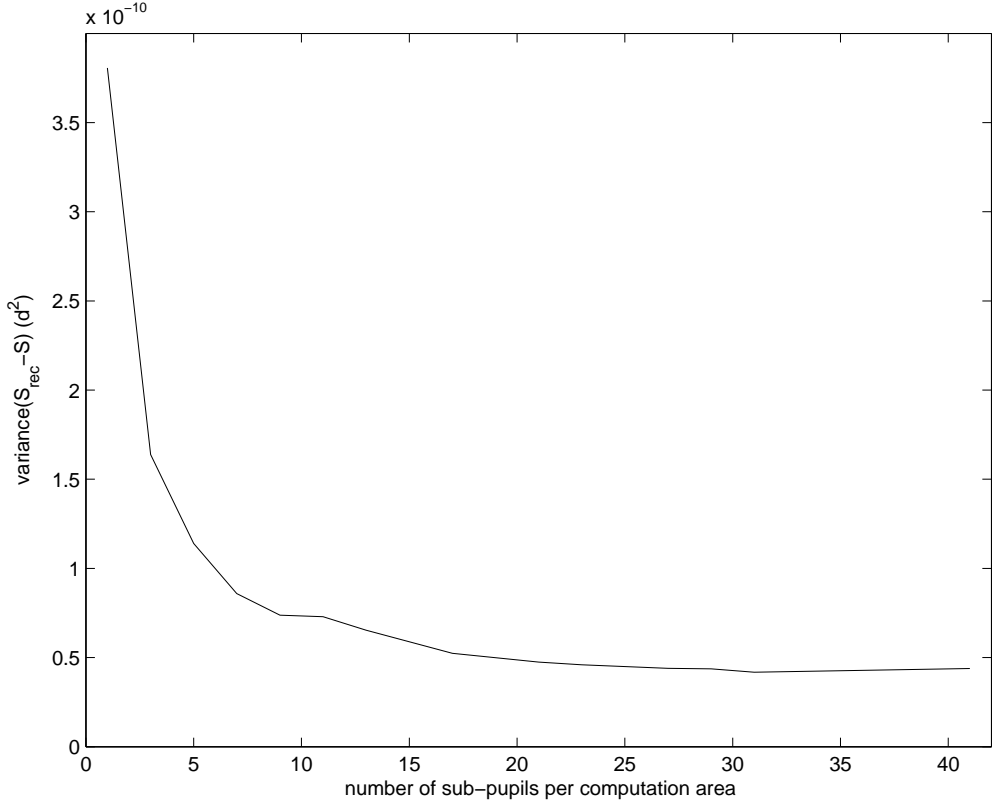}
\end{center}
\caption{\label{fig:reconstruction_error_mask} Reconstruction error
as a function of the number of selected sub-apertures.}
\end{figure}

         In order to determine the measurement sensitivity, we plot
         the variance of $(S-S_{rec})$ (where $S$ is the true pupil shift
         introduced by the PTTM) as a function of the number
         of photons received on the CCD camera as illustrated in
         figure~\ref{fig:reconstruction_error_fit}.
         For example, in our case, to be able to detect a pupil
         shift smaller than $0.1\%$ of the pupil diameter, the
         measurement error must be smaller $10^{-7} d^2$.
         We see that a minimum
         number of 7 sub-apertures located at the edge of the pupil
         (see figure~\ref{fig:reconstruction_error_mask}
         is required to fulfil the SPHERE
         specification for a reasonable number of photons received
         on the detector (about $10^5$ photons per sub-apertures).\\

         The measurement accuracy increases rapidly with the number
         of sub-apertures, but then saturates, typically when more
         than 20 sub-apertures per computation areas are
         used (figure~\ref{fig:reconstruction_error_mask}). So it shows
         that it is useless to include too many sub-apertures in
         computation areas because the ones located far from the axis
         bring more noise than signal.

         We can compare the measured variance with the analytical
         results obtained in~\ref{eq:variance_noise}.
         The reconstruction error is plotted in
         figure~\ref{fig:reconstruction_error_fit} as a function of the
         number of photons per sub-aperture. The effect of photon
         noise cannot be studied with the detector of the HASO-64
         because the CCD camera saturates before $N \simeq
         \sigma_{detec}^2$. Therefore, we only see here the influence of the detector
         noise:

         \begin{equation}
         \sigma^{2}_{I_x} \propto \frac{\sigma_{detec}^2}{2N^2}
         \end{equation}

         A very good agreement is found between the analytical
         expression and the measurements: the experimental are well
         fitted by the analytical equation.

 \section{Conclusion and perspectives}

 We have proposed a simple and efficient concept for pupil motion
tracking and conpensation during a SPHERE closed loop observation 
sequence. Analytical, simulation and experimental studies have been
conducted to validate the concept and to quantify its performances.
Each item of the erroe budget have been identified and quantified 
showing that the pupil motion sensor fullfill (with some appreciable
margins) all the SPHERE requirements.
In addition, a first experimental validation has shown the relative
simplicity of implementation and use of the device. Furthermore,
experimental results match analytical ones with a good accuracy
which is very encouraging for the future overall performance of the
pupil motion sensor en SPHERE.

\end{document}